\documentclass[12pt,letterpaper]{article}

\usepackage[ left=1in, top=1in, right=1in, bottom=1in]{geometry}
\usepackage{natbib,amsmath,float,color,bm,graphicx}
\usepackage[font={footnotesize}]{caption,subcaption}

\usepackage{array,framed}
\newcounter{algorithm}
\newenvironment{algorithm}[1][]{\refstepcounter{algorithm}\par\medskip\noindent%
   \textbf{Algorithm~\thealgorithm: #1} \rmfamily}{\medskip}

\def\zero{\mathbf{0}}

\def\NN{\mathcal{N}}

\def\IG{\mathcal{IG}}

\def\a{\mathbf{a}}  
 \def\e{\mathbf{e}}

\def\m{\mathbf{m}}

\def\v{\mathbf{v}} \def\w{\mathbf{w}} \def\x{\mathbf{x}}
\def\y{\mathbf{y}} 

  \def\C{\mathbf{C}}
  
 \def\H{\mathbf{H}} \def\I{\mathbf{I}}
 \def\K{\mathbf{K}} 
\def\M{\mathbf{M}}  
\def\P{\mathbf{P}} \def\Q{\mathbf{Q}} \def\R{\mathbf{R}}

\def\Y{\mathbf{Y}}

\newcommand{\ga}{\mbox{\boldmath $\gamma$}}

\newcommand{\bth}{\mbox{\boldmath $\theta$}}

  \def\Si{\mathbf{\Sigma}}

\DeclareMathOperator*{\argmax}{arg\,max}

\begin{document}

\title{A Bayesian adaptive ensemble Kalman filter for sequential state and parameter estimation}

\author{
Jonathan R. Stroud\thanks{McDonough School of Business, Georgetown University. jrs390@georgetown.edu} \and
Matthias Katzfuss\thanks{Department of Statistics, Texas A\&M University} \and
Christopher K. Wikle\thanks{Department of Statistics, University of Missouri}}
\maketitle

\begin{abstract}
\noindent
This paper proposes new methodology for sequential state and parameter estimation within the ensemble Kalman filter.  The method is fully Bayesian and propagates the joint posterior density of states and parameters over time.  In order to implement the method we consider two representations of the marginal posterior distribution of the parameters: a grid-based approach and a Gaussian approximation. Contrary to existing algorithms, the new method explicitly accounts for
parameter uncertainty and provides a formal way to combine information about the parameters from data at different time periods.  The method is illustrated and compared to existing approaches using simulated and real data.
\end{abstract}

\vspace{10mm}

\noindent {\em Keywords}:  Bayesian inference, filtering,
parameter estimation, spatio-temporal models, state-space models.

\newpage

%

\section{Introduction}

Data assimilation refers to sequential inference on the state of a system by combining observations with a numerical model describing the evolution of the system over time. This is an ubiquitous task in many fields, including atmospheric science, where the system is typically high-dimensional and consists of one or more spatial fields evaluated on a fine grid. From a statistical perspective, data assimilation is equivalent to filtering inference in a state-space model.  In many applications, the evolution model and other parts of the state-space model are not fully known and are instead functions of parameters. Data assimilation then requires combined inference on the (temporally varying) system state and on (temporally static) model parameters. This setting is the focus of this article.

Sequential Monte Carlo methods, also known as particle filters \cite*[][]{GordSalmSmit:93,PittShep:99,DoucFreiGord:01}, are widely used for sequential estimation in general state-space models.  Although there is an enormous literature on pure state estimation, there are fewer papers that consider sequential estimation of both states and parameters.  The existing references include \cite*{Kita:98}, who proposed augmenting the state vector to include the static parameter and then estimating the augmented state using the particle filter. \cite{LiuWest:01} proposed another state augmentation approach that uses kernel density estimation of the parameter distribution within an auxiliary particle filter \citep{PittShep:99} framework.  \cite{Stor:02} suggested analytical updating of sufficient statistics, but this approach only applies to parameters with conjugate priors.  \cite*{AndrDouc:05} proposed recursive and batch MLE methods. These methods have all been shown to work well in nonlinear and non-Gaussian models, when the state dimension is fairly small, say less than 10 dimensions. However, these particle filters rely on reweighting or resampling of particles, which results in filter collapse when the state dimension is high \citep[e.g.,][]{SnydBengBick:08}. Hence, these particle-filter-based methods are not suited for the high-dimensional systems of interest here.

The ensemble Kalman Filter \citep[EnKF;][]{Even:94} is a sequential Monte Carlo algorithm designed for combining high-dimensional space-time models with observations. Several reviews of and tutorials on the EnKF are available \citep[e.g.,][]{Wikle2007,Even:09,Katzfuss2015b,HoutZhan:16}.
 While the EnKF is closely related to the Kalman filter \citep[KF;][]{Kalm:60}, it handles nonlinearities in a more flexible manner than analytic linearization schemes such as the extended Kalman filter \citep[e.g.,][Ch.~5]{GrewAndr:93}.  Although much work has been done to improve EnKF estimation of state variables, little work has focused on estimation of model parameters.  \cite{Ande:01} proposed adding the unknown parameters to the state vector and updating the augmented state using a standard EnKF scheme, but this state-augmentation approach does not work well for parameters that exhibit small (linear) correlation with the state vector. For example, \cite{StroBeng:07} and \citet{DelsYang:10} show that the augmentation method fails for variance parameters.

Here, we consider parameter inference in the EnKF based on the likelihood, which is the distribution or density of the observed data conditional on the parameters, viewed as a function of the parameters. 
Offline maximum likelihood (ML) estimation in the EnKF framework has been considered using Newton-Raphson \citep{DelsYang:10,StroSteiLesh:10}, as well as grid-based \citep{UenoHiguKagi:10} and expectation-maximization \citep{TandPuliLott:15} optimization techniques. ML estimation of parameters from an online perspective was considered by \cite{MitcHout:00}, whose method combines ML estimates at each time point in a statistically inconsistent way (see Section \ref{sec:likapprox} later); by \cite{UenoNaka:14}, who estimate parameters in the noise covariance matrix via online expectation-maximization algorithm; and by \cite{De:14}, who proposed a method for sequentially updating the unknown parameters at each time point to find a stochastic approximation to the ML estimator in stationary systems.

The likelihood can also be used to conduct Bayesian inference on the parameters. \cite{StroBeng:07} provide a Bayesian method for parameter inference within the EnKF, but their approach is limited to a scalar variance parameter describing the magnitude of additive evolution-model error.
\cite{FreiKuns:12} propose Bayesian inference on parameters by combining an EnKF for state inference with a particle filter to approximate the parameter distribution, but their focus is on temporally varying observation error covariance parameters. \cite{BranCosmTest:10} find the maximum a posteriori (MAP) estimators of temporally varying parameters, while \cite{UenoNaka:16} focus on MAP estimation for parameters in the noise covariance matrix with temporal smoothing via online EM.

Here, we propose a fully Bayesian method for sequential (i.e., online) inference on states and parameters within the EnKF framework. Our algorithms are designed to be applicable to temporally static parameters in nonlinear, high-dimensional systems. Unlike some of the other approaches \citep[e.g.][]{MitcHout:00}, our method combines information about the parameters from data at different time points in a formal way using Bayesian updating. In contrast to the ML and MAP approaches discussed above, we quantify uncertainty in the parameters through analytic propagation of the entire filtering distribution of the parameters. Further, our approach is suitable for static parameters in various parts of the state-space model, including in the evolution-error and noise covariance matrices. To implement our algorithm, we propose two approximate methods: one based on a parameter grid and 
another based on a normal approximation to the parameter distribution.

Note that there is also an extensive literature on estimation of specific tuning parameters in the EnKF, such as inflation and localization parameters \citep[e.g.][]{WangBish:03,Ande:07a,Ande:07b}. Here we focus instead on inference on general parameters that explicitly appear in the statistical model, and we regard the tuning parameters as known.

The remainder of the paper is outlined as follows. In Section \ref{sec:notation}, we introduce the state-space model under consideration. In Section \ref{sec:methodology}, we motivate and introduce our proposed methodology. In Section \ref{sec:applications}, we present numerical comparisons of our methods to existing approaches using simulated and real data. We conclude in Section \ref{sec:conclusions}.

\section{Additive Gaussian state-space models \label{sec:notation}}

Let $\y_t$ denote the $m_t \times 1$ observation vector and $\x_t$ the $n \times 1$ state vector.  We consider the following class of additive Gaussian state-space models:
\begin{align}
&  && \mbox{Observation:} &&
\y_t \;=\; \H_t(\bth)\x_t + \v_t,    &&\v_t \sim \NN(\zero,\R_t(\bth)) && &&
\label{eq:obs}\\
&  && \mbox{Evolution:} &&
 \x_t \;=\; \mathcal{M}_t(\x_{t-1};\ga) + \w_t, &&\w_t \sim \NN(\zero,\Q_t(\bth)), && &&
\label{eq:evo}
\end{align}
for $t=1,2,\ldots$, 
where the observation matrix $\H_t$ and the covariance matrices $\R_t$ and $\Q_t$ may depend on a vector of unknown parameters, $\bth$, and the possibly nonlinear evolution operator $\mathcal{M}_t(\cdot)$ may depend on a separate set of parameters, $\ga$. For now, we will focus on inference for $\bth$ and consider $\ga$ to be fixed and known (and so we simply write $\mathcal{M}_t(\x_{t-1})$), but we will describe in Section \ref{sec:merge} how our algorithm for inference on $\bth$ can be combined with a state-augmentation approach to perform inference on $\ga$.
The model is completed with a prior on the initial state, $p(\x_0|\bth) = \NN(\a_0(\bth), \P_0(\bth))$, and a prior distribution of the parameters, $p(\bth)$. In applications where the relationship between the observations and the state is nonlinear, we take $\H_t$ in \eqref{eq:obs} to be the linearization of the nonlinear mapping.


\section{Sequential Bayesian inference on state and parameters \label{sec:methodology}}

The Bayesian filtering problem requires computing the joint posterior distribution $p(\x_t,\bth|\Y_t)$ of the current state and the parameters at each time $t=1,\ldots,T$, where $\Y_t = \{\Y_0, \y_1, \ldots, \y_t\}$ denotes the information available at time $t$, and $\Y_0$ is the initial information. This joint posterior encodes all available information about the states and parameters contained in the data, and it is typically summarized through marginal distributions, posterior means, standard deviations, or credible intervals. As we will see, Bayesian inference has two advantages over frequentist or more ad-hoc methods: It allows accounting for parameter uncertainty, and information about the parameters can be naturally combined over time following a consistent probabilistic framework.

Except in very special cases \cite[see][]{StroBeng:07}, the joint posterior distribution is unavailable in closed form, so Monte Carlo methods must be used to approximate the distribution.  In what follows, we propose a method for combined state and parameter estimation that scales to high-dimensional states.

Our approach relies on the decomposition of the joint posterior distribution of the state and parameters into two terms: the conditional posterior distribution for the states given the parameters, and the marginal posterior distribution for the parameters:
\begin{equation}
\label{eq:posteriordecomposition}
p(\x_t,\bth|\Y_t)  = p(\x_t|\bth,\Y_t) p(\bth|\Y_t).
\end{equation}
In the following subsections, we describe how $p(\x_t|\bth,\Y_t)$ can be obtained via the EnKF (Section \ref{sec:enkfstate}), we examine the marginal parameter posterior $p(\bth|\Y_t)$ (Section \ref{sec:parameterposterior}), we propose two approximation methods for $p(\bth|\Y_t)$ (Section \ref{sec:parameterapprox}), and finally, we will describe the full algorithm that combines these ideas and results (Section \ref{sec:algorithm}).

\subsection{EnKF for state inference \label{sec:enkfstate}}

The first term on the right-hand side of \eqref{eq:posteriordecomposition} is the filtering distribution of the state given the parameters. Since our algorithm must be implemented sequentially, it is useful to write this distribution in recursive form:
\begin{equation}
p(\x_t|\bth,\Y_t) ~ \propto ~ p(\y_t|\x_t,\bth)  
\; \int p(\x_t|\x_{t-1},\bth) \; p(\x_{t-1}|\bth,\Y_{t-1}) \; d\x_{t-1},
\label{eqn:cond}
\end{equation}
that is, the observation density times the state forecast density, which is defined by the integral.  In a linear Gaussian model, these recursions can be computed analytically using the Kalman filter (provided the dimension of the state is not excessively large).

Here, we are interested in high-dimensional systems with possibly nonlinear evolution (see Section \ref{sec:notation}), for which we instead employ an EnKF. Assume we have an ensemble of states, $\{\x_{t-1}^{(i)}\}_{i=1}^N$, representing the filtering distribution at time $t-1$. The EnKF then propagates each state vector forward, $\x_t^{pi}= \mathcal{M}(\x_{t-1}^i)$, $i=1,\ldots,N$, and estimates the covariance matrix from the prior ensemble. In most applications we have $n \gg N$, and some form of regularization of this covariance matrix is necessary. Denoting the prior ensemble mean as $\hat{\a}_t^p = \frac{1}{N} \sum_{i=1}^N \x_t^{pi}$, we assume here that 
\begin{equation}
\label{eq:priorcov}
\hat{\P}_t^p =   \bm{\rho} \circ \textstyle\big(\frac{1}{N-1} \sum_{i=1}^N (\x_t^{pi}-\hat{\a}_t^p)(\x_t^{pi}-\hat{\a}_t^p)'\big)
\end{equation}
is given by an elementwise product of the empirical covariance matrix with a sparse tapering correlation matrix  
$\bm{\rho}$ \citep[e.g.,][]{HoutMitc:98,Ande:07b,FurrBeng:07}. The estimated Kalman gain is a function of $\bth$:
\begin{equation}
\label{eq:gain}
\hat{\K}_t(\bth)=\H_t(\bth)\hat{\P}_t^{f}(\bth)\H_t(\bth)'\hat\Si_t(\bth)^{-1},
\end{equation}
where $\hat{\P}_t^{f}(\bth) = \hat{\P}_t^p + \Q_t(\bth)$ and 
\begin{equation}
\label{eq:innovcov}
\hat\Si_t(\bth) = \H_t(\bth)\hat{\P}_t^{f}(\bth)\H_t(\bth)'+\R_t(\bth).
\end{equation}
Then, after generating the forecast ensemble by setting $\x_t^{fi} = \x_t^{pi}+ \w_t^i$, where $\w_t^i \sim \NN(\zero,\Q(\bth))$, $i=1,\ldots,N$, and simultating observation errors as $\v_t^i \sim \NN(\zero,\R_t(\bth))$, $i=1,\ldots,N$, we can obtain the posterior ensemble at time $t$ based on parameter value $\bth$ using the analysis scheme of \cite{BurgVanlEven:98}:
\begin{equation*}
\x_t^i ~=~ \x_t^{fi}+ \hat{\K}_t(\bth)\big(\y_t+\v_t^i+\H_t(\bth)\x_t^{fi}\big).
\end{equation*}

\subsection{The marginal posterior of parameters \label{sec:parameterposterior}}

The second term on the right-hand side of \eqref{eq:posteriordecomposition} is the marginal posterior for the
parameters. It can be written recursively as
\begin{equation}
p(\bth|\Y_t) ~ \propto   ~ p(\y_t|\bth,\Y_{t-1}) ~ p(\bth|\Y_{t-1}).
\label{eqn:marg}
\end{equation}
The above formula is crucial as it defines a recursion for the
parameter distribution over time.  Note that the first term on the
right side is the likelihood at time $t$ and the second term is the ``previous'' 
parameter posterior at time $t-1$. 
 
Note that, under a flat initial prior for the parameters (i.e., $p(\bth) \propto 1$), the marginal
posterior for $\bth$ in \eqref{eqn:marg} is exactly proportional to the cumulative likelihood
$L_t(\bth)=\prod_{j=1}^t p(\y_j|\bth,\Y_{j-1})$ that is often considered in frequentist (i.e., non-Bayesian) inference.

\subsubsection{Accumulation of evidence over time \label{sec:accumulation}}

The recursion in \eqref{eqn:marg} provides a natural
way to propagate information about the parameter over time, as opposed
to the ad hoc methods used by \cite{Dee:95} and \cite{MitcHout:00}. To illustrate this, we replicated a static-model example presented in \cite{MitcHout:00}.  The model assumes that (scalar) observations $y_t \sim \NN(0,2+\alpha)$ are generated independently from a Gaussian distribution with mean zero and variance $(2+\alpha)$, where $\alpha$ is an unknown variance parameter, and the goal is to estimate $\alpha$ sequentially as new data arrive. \cite{MitcHout:00} considered the cumulative mean and median of the single-stage ML estimates $\hat\alpha_t = \arg \max_\alpha p(y_t|\alpha)$, where $p(y_t|\alpha)$ is the likelihood considering only the data at time $t$. Because $\alpha$ is a variance parameter and thus must be nonnegative, the estimates are given by $\hat\alpha_t = \max(0,y_t^2-2)$.

\begin{figure}
\centering
\includegraphics[width=.7\textwidth]{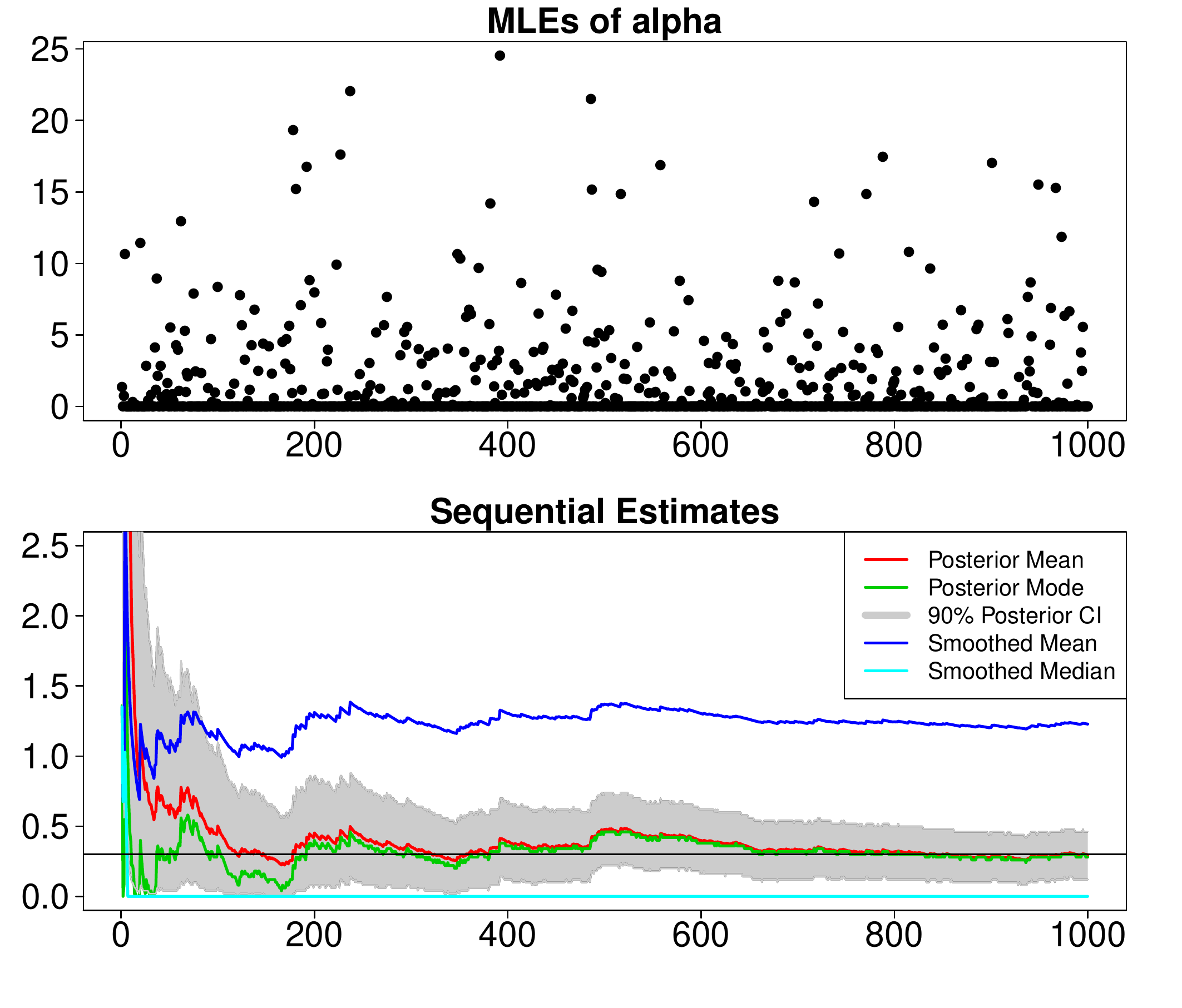}
\caption{For the static model in Section \ref{sec:accumulation}, individual maximum likelihood estimates (top panel) and cumulative inference (bottom) on the parameter $\alpha$ over time. The true parameter value is $\alpha_*=0.3$.  In the bottom panel, the blue lines represent the cumulative mean and median of the individual estimates as proposed by \cite{MitcHout:00}, while the red and green lines and the gray band represent summaries of the posterior distribution of $\alpha$. \label{fig:MHcomp}}
\end{figure}

The results of a simulation using a true parameter value $\alpha_*=0.3$ are shown in Figure \ref{fig:MHcomp}. The majority of the individual estimates $\hat\alpha_t$ are equal to zero, and the distribution of the estimates is heavily right-skewed. Hence, as in \cite{MitcHout:00}, we find an upward bias in the cumulative mean with respect to the true value $\alpha_*$, while the cumulative median is zero. Moreover, after 10,000 observations the estimates did not converge to $\alpha_*$, indicating statistical inconsistency of the estimators. In contrast, the posterior distribution of $\alpha$ from \eqref{eqn:marg} can be seen to become more concentrated over time and to converge to the true value of $\alpha_*$. Corresponding point estimates, such as the posterior mode and posterior mean, do converge to the true value. 

Thus, it is desirable to use the recursive expression for the posterior distribution of the parameters from \eqref{eqn:marg} for rigorously combining information about the parameters from data at different time points.

\subsubsection{Feasible likelihood approximation for high-dimensional states \label{sec:likapprox}}

\begin{figure}
	\begin{subfigure}{.32\textwidth}
	\centering
	\includegraphics[width =1\linewidth]{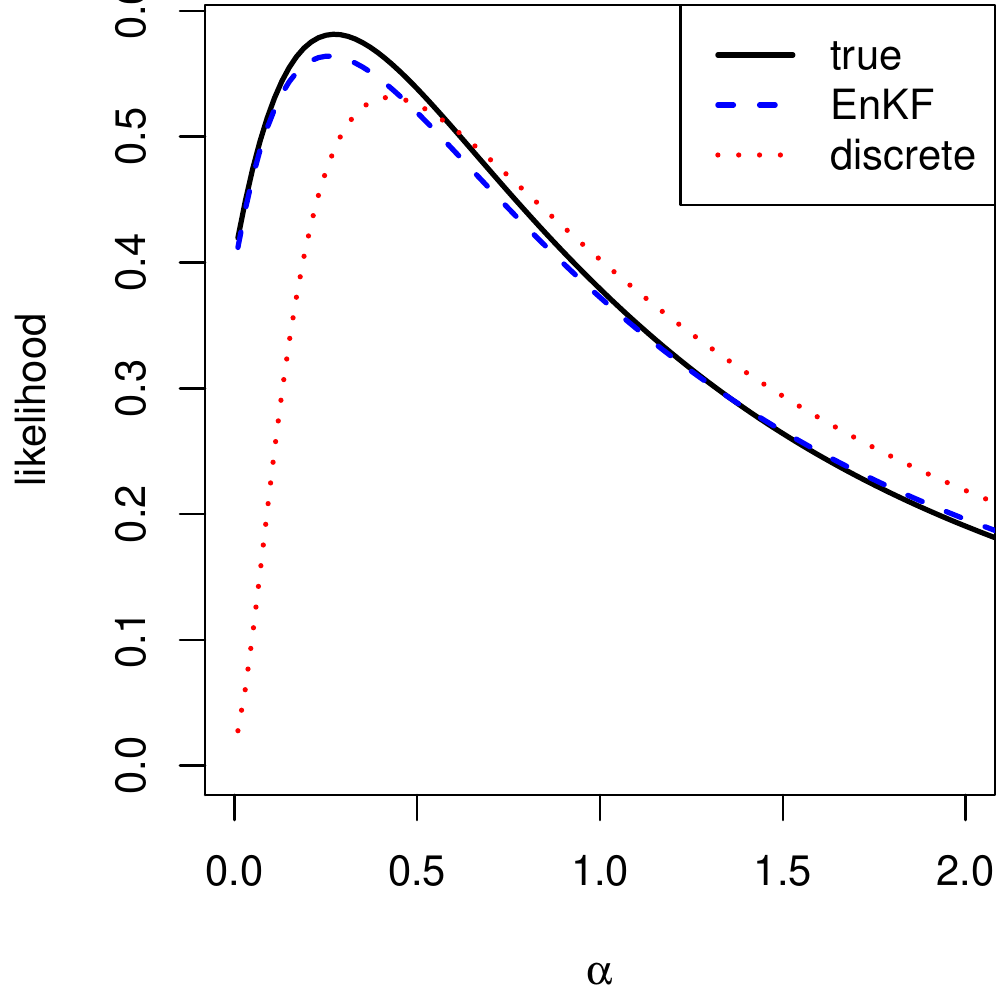}
	\caption{$n=5$}
	\end{subfigure}%
\hfill
	\begin{subfigure}{.32\textwidth}
	\centering
	\includegraphics[width =1\linewidth]{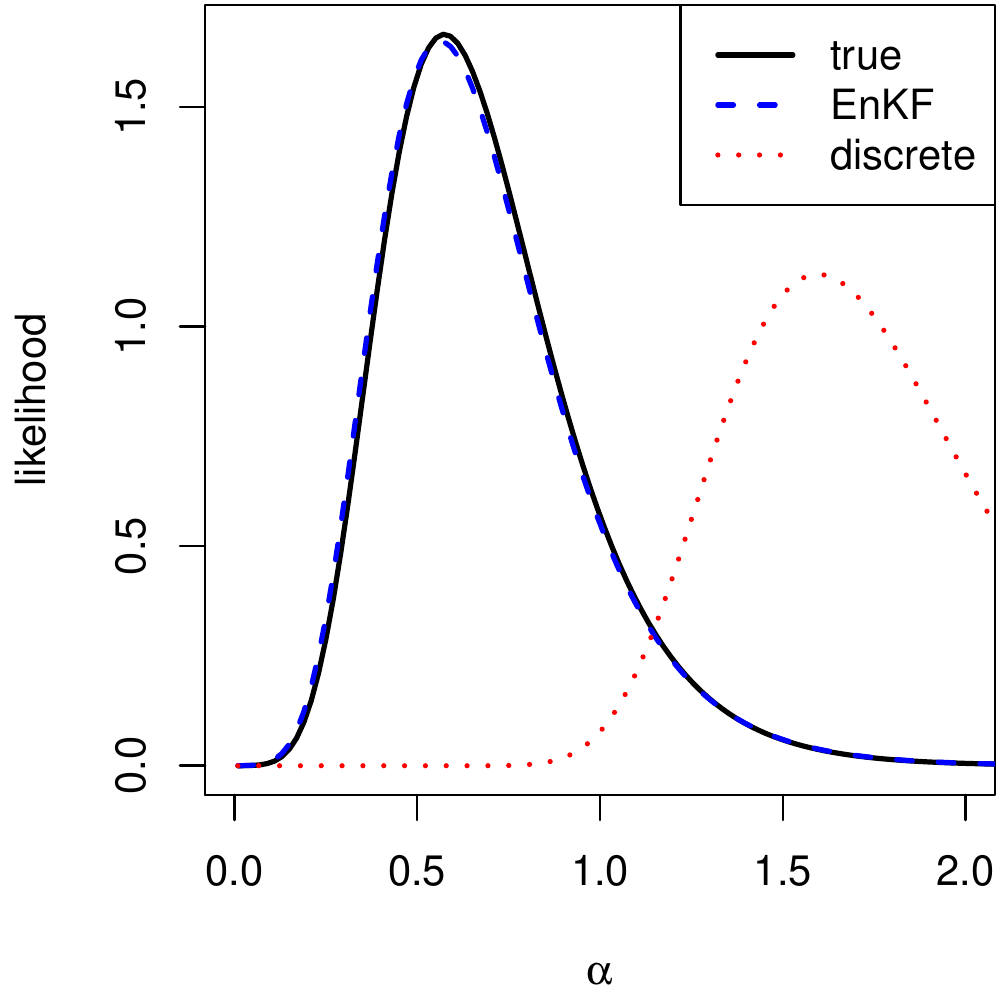}
	\caption{$n=50$}
	\end{subfigure}%
\hfill
	\begin{subfigure}{.32\textwidth}
	\centering
	\includegraphics[width =1\linewidth]{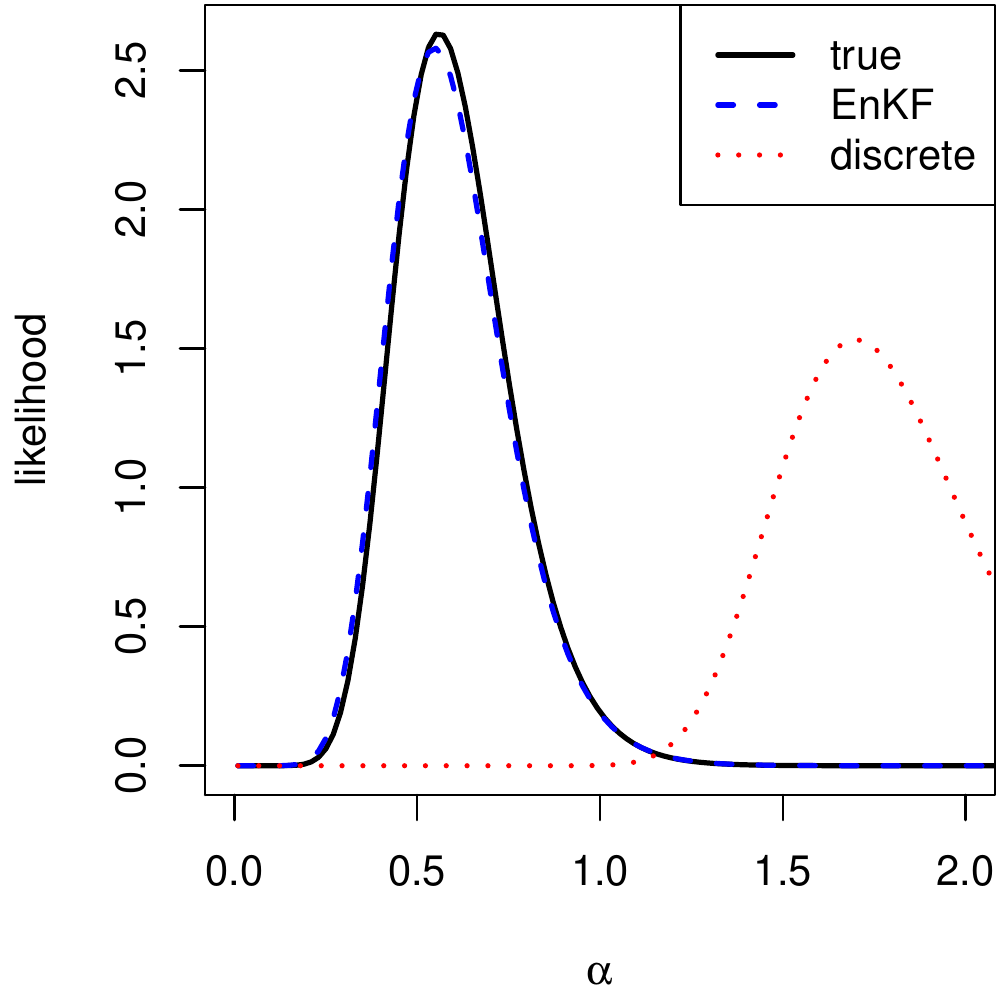}
	\caption{$n=100$}
	\end{subfigure}
\caption{Comparison of the likelihoods discussed in Section \ref{sec:likapprox} in a simple simulated example. Data are simulated at a single time point from prior covariance matrix $\P_t^p$, with $\Q = \alpha\I_n$, $\H=\I_n$, $\R=0.1 \I_n$. The true $\alpha$ is 0.5. The observations are on a one-dimensional spatial domain at locations $1,2,\ldots,n$, and $\P_t^p$ is based on an exponential covariance function with range parameter 3. We use the same $N=50$ draws from the forecast distribution for the EnKF and discrete approximations, and a Wendland taper with range 12 for the EnKF. The likelihoods are normalized to integrate to 1, and can hence be viewed as posterior distributions for $\alpha$ under a uniform (flat) prior.}
\label{fig:likcomparison}
\end{figure}

For high-dimensional models, evaluation of the likelihood $p(\y_t|\bth,\Y_{t-1})$ in \eqref{eqn:marg} is computationally infeasible. Given that we use an ensemble representation for the state distributions in Section \ref{sec:enkfstate} above, it is natural to use the same ensemble representation in order to approximate the likelihood. Specifically, given that the filtering distribution at time $t-1$ is a discrete distribution with equal weights at the filtering ensemble $\{\x_{t-1}^{(i)}\}_{i=1}^N$, an approximation of the likelihood (called the ``discrete'' likelihood approximation here) is given by
\[
p(\y_t|\bth,\Y_{t-1}) = \textstyle\frac{1}{N} \sum_{i=1}^N \NN\big(\y_t\big| \H_t(\bth)\x_t^{pi},\H_t(\bth)\Q_t(\bth)\H_t(\bth)'+\R_t(\bth)\big)
\]
However, as illustrated in Figure \ref{fig:likcomparison}, 
this approximation can break down when the data are informative (i.e., when $m_t$ and the signal-to-noise ratio are large relative to $N$).
Instead, we employ a likelihood approximation based on the EnKF, which approximates the forecast distribution by a multivariate Gaussian distribution \citep{MitcHout:00}. This EnKF likelihood approximation is given by
\begin{equation}
p(\y_t|\bth,\Y_{t-1}) ~ \propto ~ |\hat\Si_t(\bth)|^{-\frac{1}{2}}
\exp\left\{-(1/2)\cdot\hat\e_t(\bth)'\hat\Si_t(\bth)^{-1}\hat\e_t(\bth)\right\},
\label{eqn:likelihood}
\end{equation}
where the innovation covariance $\hat\Si_t(\bth)$ is given in \eqref{eq:innovcov} above, and the innovation is given by $\hat{\e}_t(\bth) =\y_t-\H_t(\bth)\hat{\a}_t^p$.

\subsection{Approximations of the parameter posterior \label{sec:parameterapprox}}

 Note that the likelihood $p(\y_t|\bth,\Y_{t-1})$ in \eqref{eqn:marg} is a complicated
nonlinear function of $\bth$, as it arises in both the determinant
and inverse of an $n\times n$ matrix.  Thus, the marginal posterior
distribution $p(\bth|\Y_t)$ in (\ref{eqn:marg}) is typically
unavailable in closed form.

To perform parameter learning in the Bayesian framework, we need a
representation of the parameter distribution which allows for
recursive updating.  In what follows, we consider two representations
of the parameter distribution: a discrete (gridded) distribution and a Gaussian
approximation.  The first provides an exact recursive updating method
on a discretized parameter space, while the second method provides a
approximate method over the full parameter space.

\subsubsection{Grid-based representation of $p(\bth|\Y_t)$ \label{sec:gridapprox}}

This approach treats the parameter space as discrete.  The parameter
distribution is specified by a set of points $\{\bth_1, \ldots,
\bth_K\}$ and associated probability weights $\{\pi_{t,1}, \ldots,
\pi_{t,K}\}$ with normalization constraint $\sum_{k=1}^K\pi_{t,k}=1$.
The discrete parameter distribution is defined by
\begin{equation}
p_d(\bth|\Y_t) ~=~ \sum_{k=1}^K \delta_{\theta_k}(\bth) \, \pi_{t,k}, 
\label{eqn:disc_post}
\end{equation}
where $\delta(\cdot)$ is the Kronecker delta function.  With this representation
for the prior, the posterior is given as the product of the prior and 
likelihood; that is,
\begin{equation}
p_d(\bth|\Y_t) ~ \propto ~ p(\y_t|\bth,\Y_{t-1})\;p_d(\bth|\Y_{t-1}).
\label{eqn:disc_update}
\end{equation}
We assume that the parameter grid is fixed over time.  The updating
formula in \eqref{eqn:marg} reduces to a recursion on the weights:
$\pi_{t,k}\propto p(\y_t|\bth_k,\Y_{t-1}) \pi_{t-1,k}$, for
$k=1,\ldots,K$, where the weights $\pi_{t,k}$ are normalized to sum to
1.  Therefore, the computational cost of the update is $K$ likelihood
evaluations, one for each gridpoint.

While the discrete approach is conceptually appealing, it has some limitations.  First, the method does not extend beyond a few parameters since the computational cost grows exponentially in the dimension of the parameter space.  Second, the grid of parameter values must be specified a priori, and is not adaptive over time.  Thus, as the posterior becomes more concentrated over time, the posterior distribution eventually concentrates on a single grid point. This implies falsely that there is no posterior uncertainty about $\bth$.  To alleviate these problems, we next consider another method based on a Gaussian approximation.

\subsubsection{Normal approximation to $p(\bth|\Y_t)$ \label{sec:normalapprox}}

Here, the parameter distribution at each time $t$ is approximated by a normal distribution with mean $\m_t$ and covariance matrix $\C_t$.  The posterior density is then given by
\begin{equation}
p_n(\bth|\Y_t) ~ \propto 
\exp\left\{-\frac{1}{2}(\bth-\m_t)'\C_t^{-1}(\bth-\m_t)\right\}.
\label{eqn:norm_post}
\end{equation}
The updating recursions for the posterior moments are then derived as follows. Assume the parameter distribution at time $t-1$ is normal with mean $\m_{t-1}$ and covariance $\C_{t-1}$.  The posterior is proportional to the product of likelihood and prior; that is, we take
\begin{equation}
p_n(\bth|\Y_t) ~ \approx ~ p(\y_t|\bth,\Y_{t-1})\;p_n(\bth|\Y_{t-1}).
\label{eqn:norm_update}
\end{equation}
Since this posterior is not of a recognizable form, a normal approximation is used.  Let $\exp\{\ell(\bth)\}$ denote the right-side of \eqref{eqn:norm_update}.  This approximation is defined by a second-order expansion of $\ell(\bth)$ at the mode.  The posterior mean and covariance are defined by
\begin{equation}
\m_t\,=\,\argmax_{\bth} \, \ell(\bth) ~~~\mbox{and}~~~
\C_t\,=\,-\left[
\frac{\partial^2 \ell(\bth)}
{\partial\theta_i \partial\theta_j}\right]^{-1}_{\bth=\m_t}.
\label{eqn:norm_moments}
\end{equation}
A numerical optimization scheme is used to obtain the posterior mean.  Since $\ell(\bth)$ is not concave, a global optimum is not guaranteed; however, in our examples we find that the optimizations are quite reliable.

\subsection{Combined state and parameter learning in the EnKF \label{sec:algorithm}}

Given the developments above, an ensemble-based algorithm is proposed to generate a sample from the joint posterior distribution of the state and parameters in \eqref{eq:posteriordecomposition} at each time point. At each $t$, we have an analytical (discrete or normal) representation of the parameter distribution, $\hat{p}(\bth|\Y_t)$, along with an ensemble of states and parameters $\{(\x_t^{(i)},\bth^{(i)})\}_{i=1}^N$ from $p(\x_t,\bth|\Y_t)$.

Our approach is closely related to that of \cite{MitcHout:00}, but it includes steps to update and simulate from the parameter distribution rather than obtaining $\bth$ through maximum likelihood. Our approach naturally quantifies uncertainty in the parameters, and takes this uncertainty into account when obtaining the filtering ensemble of the state.

In the following algorithm, the superscripts $p$ and $f$ refer to the predictive and forecast distributions, the superscript $i$ is the ensemble index, and $N$ is the ensemble size.  


\begin{framed}
\begin{algorithm}
\label{alg:enkflearning} 
The algorithm is initialized by drawing from the initial
prior: $\bth^i \sim p(\bth)$ and $\x^i \sim \NN(\a_0,\P_0)$ for $i=1,\ldots,N$.  
Each assimilation cycle $t=1,2,\ldots$ then proceeds as follows:  
\begin{enumerate}
\item Propagate each state vector forward:
\begin{equation}
\x_t^{pi}= \mathcal{M}(\x_{t-1}^i), ~~~ i=1,\ldots,N.
\end{equation}
\item Approximate the likelihood function using the prior ensemble as in \eqref{eqn:likelihood} by
\begin{equation*}
\hat{p}(\y_t|\bth,\Y_{t-1}) ~ \propto ~ |\hat{\Si}_t(\bth)|^{-\frac{1}{2}}
\exp\left\{-(1/2)\,\hat{\e}_t(\bth)'\hat{\Si}_t(\bth)^{-1}\hat{\e}_t(\bth)\right\}.
\label{eqn:like_appx}
\end{equation*}
\item Update the analytical parameter distribution using the grid-based (Section \ref{sec:gridapprox}) or the normal (Section \ref{sec:normalapprox}) approximation:
\begin{equation*}
\hat{p}(\bth|\Y_t) ~ \propto ~ \hat{p}(\y_t|\bth,\Y_{t-1})\hat{p}(\bth|\Y_{t-1}).
\end{equation*}
\item Draw parameters from the updated posterior distribution:
\begin{equation*}
\bth^i ~ \sim ~ \hat{p}(\bth|\Y_t), ~~ i=1,\ldots,N.
\end{equation*}
\item Generate the forecast ensemble by setting 
\begin{equation*}
\x_t^{fi} ~=~ \x_t^{pi}+ \w_t^i,
~~~\mbox{where}~~~ \w_t^i \sim \NN(\zero,\Q(\bth^i)), ~~ i=1,\ldots,N.
\end{equation*} 
\item Draw a posterior ensemble using the analysis scheme of 
\cite{BurgVanlEven:98}:
\begin{equation*}
\x_t^i ~=~ \x_t^{fi}+ \hat{\K}_t(\bth^i)(\y_t+\v_t^i+\H_t(\bth)\x_t^{fi}), 
~~~ \mbox{where} ~~~
\v_t^i \sim \NN(\zero,\R_t(\bth^i)), ~~ i=1,\ldots,N,
\end{equation*}
and the estimated Kalman gain $\hat{\K}_t(\bth)$ is given in \eqref{eq:gain}.
\end{enumerate}
\end{algorithm}
\end{framed}

Depending on which approximation method is used in Step 3 of this algorithm, we refer to it as EnKF-Grid or EnKF-Normal.

\subsubsection{Merging our method with existing approaches \label{sec:merge}}

Note that our algorithm works best when the number of unknown parameters in $\bth$ is small. Hence, we recommend combining our algorithm with other approaches as much as possible. 

For example, it can be combined with state augmentation \citep{Ande:01}, which works well for parameters ($\ga$, say) that have a strong correlation with the state (e.g., parameters in $\mathcal{M}$ as introduced in \eqref{eq:evo}). In the algorithm above, this means replacing the state $\x_t$ by the augmented state $(\x_t',\ga')'$, and $\H_t$ by the matrix $(\H_t, \zero)$. The transition of the parameters $\ga$ is typically assumed to be constant, although it is also possible to treat $\ga$ as a time-varying parameter $\ga_t$ with small artificial evolution noise \citep[e.g.,][]{Kita:98,LiuWest:01}.   
Equivalently, we could use covariance inflation for the parameters, which has a similar effect.

The EnKF-Grid method can also be combined with the approach of \citet{StroBeng:07} to make inference on a scalar multiplicative parameter that appears in both $\Q_t$ and $\R_t$. If this parameter has an inverse-gamma prior distribution, its marginal posterior distribution is also inverse gamma and available in closed form. \cite{StroBeng:07} provide an EnKF algorithm to update the hyperparameters of the inverse-gamma distribution at each time point and sample from the joint filtering distribution of the state vector and the scalar parameter.

\section{Numerical comparison and applications \label{sec:applications}}


\subsection{Linear evolution \label{sec:linapp}}

We first consider a linear dynamic spatio-temporal model from \cite{XuWikl:07}.  
The model is a vector autoregression plus noise where the state 
vector $\x_t=(x_{t1},\ldots,x_{tn})'$ corresponds to $n$ equally-spaced locations $\{1,2,3...,n\}$
along a spatial transect.  Following the notation in (1)--(2), the evolution mean 
function is linear, $\mathcal{M}_t(\x_{t-1}) = \M\x_{t-1}$, where the propagator 
matrix is tridiagonal with parameters $\ga=(\gamma_1,\gamma_2,\gamma_3)$:
$$\M(\ga) = \left(
\begin{tabular}{cccc}
$\gamma_1$ & $\gamma_2$ &            &     0        \\
$\gamma_3$ & $\gamma_1$ & $\ddots$   &              \\
           &  $\ddots$  & $\ddots$   & $\gamma_2$ \\
     0     &            & $\gamma_3$ & $\gamma_1$ 
\end{tabular}\right).
$$ 
The evolution errors are spatially correlated with covariance $\Q(\bth) = \sigma^2_\eta \C(\tau)$, where $\C(\tau)$ is defined by the exponential correlation function $c(d;\tau)=\,\exp(-\tau d)$, and $d$ is the distance between locations. The initial state distribution is given by $p(\x_0|\bth) = \NN(\zero,\sigma^2_\epsilon\I)$. For the data model, we assume that observations are taken at each location, $\y_t=(y_{t1},\ldots,y_{tn})'$, and the observation matrix and error covariance matrix are given by $\H=\I$ and $\R=\sigma^2_\epsilon\I$. The signal-to-noise ratio is denoted by $\beta = \sigma^2_\eta/\sigma^2_\epsilon$.

We consider two relatively low-dimensional examples here, which (along with the assumption of linear evolution) allows us to compute the true posterior distribution of $\bth$ at each time using the Markov chain Monte Carlo (MCMC) procedure of \citet{CartKohn:94}, and to compare ours and other approaches to the true posterior distribution.


\subsubsection{Simulation \label{sec:linsim}}

First, we simulated observations from the true model with dimensions $n=m=20$ for $T=100$ time points. The true parameters were taken to be $\ga=(.3,.6,.1)'$, $\beta=5$, $\tau=1$, and $\sigma^2_\epsilon=1$. For this simulation, we assumed that $\ga$ and $\sigma^2_\epsilon$ were known, so that $\bth=(\beta,\tau)'$ were the unknown parameters with independent prior distributions $\beta \sim \NN^+(5,10)$ and $\tau \sim \NN^+(2,.16)$, where $\NN^+$ denotes a truncated normal distribution on the positive real line.

\begin{figure}
\centering
\includegraphics[width=.9\textwidth]{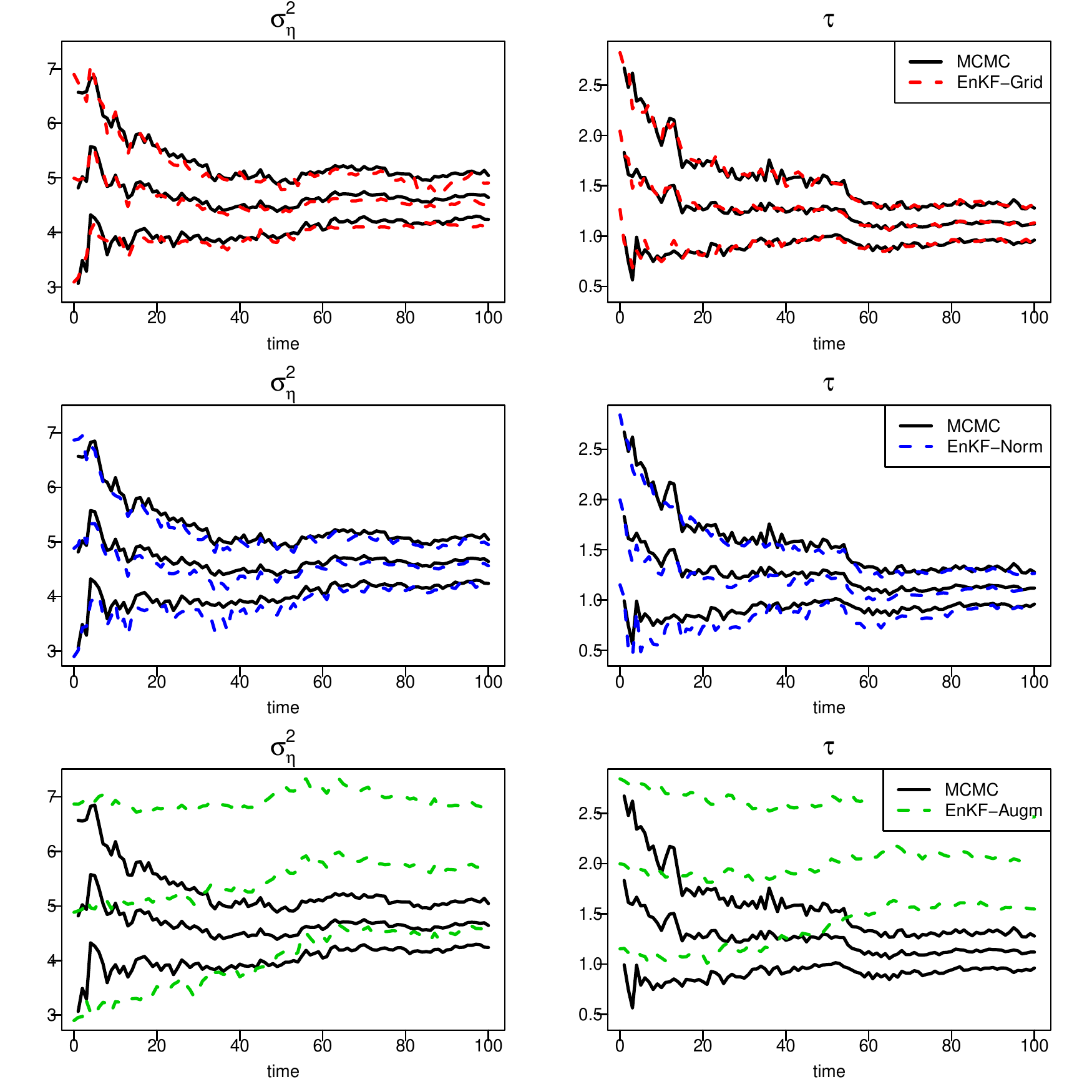}
\caption{For the simulated data using a linear state-space model described in Section \ref{sec:linsim}, true posterior distributions (filtered mean and 95\% bands) over time of the parameters $\sigma^2_\eta$ and $\tau$ (in black) and the corresponding approximations using the EnKF-Grid (top), EnKF-Norm (middle), and EnKF with state augmentation (bottom) 
\label{fig:simlinear}}
\end{figure}

We obtained the posterior distribution of the parameters $\bth$ for each time $t=1,\ldots,T$ using Algorithm \ref{alg:enkflearning} with $N=100$ ensemble members and no tapering or covariance inflation. The results are shown in Figure \ref{fig:simlinear}. As we can see, the posterior distributions as approximated by our EnKF-Norm and EnKF-Grid procedures from Algorithm \ref{alg:enkflearning} are very close to the true posterior distribution obtained via MCMC, and they seem to converge to the true values of $\sigma^2_\eta = \beta \sigma^2_\epsilon = 5$ and $\tau = 1$. This is in contrast to the approximation of the posterior obtained by state augmentation.  The flat bands for both parameters indicate that the augmentation approach does not work in this case, likely because the relationship between the covariance parameters and the observations is not linear.

\subsubsection{Cloud data \label{sec:cloud}}

Next, we apply the proposed methods to the cloud motion data of \cite{Wikl:02}.  The data are cloud intensities at $n=60$ equally-spaced locations along a transect at $T=80$ time periods.  \cite{Wikl:02} used non-Gaussian spatio-temporal kernel models to analyze the data.  Here, since the original data, $z_{ti}$,  are counts with a large number of zeros, we work with the transformed observations, $y_{ti}=\log (1+z_{ti})$. Using again the model of \cite{XuWikl:07} described above, we now treat all parameters $\bth = (\ga',\beta,\tau,\sigma^2_\epsilon)'$ as unknown with the following prior distributions:
$\ga | \sigma^2_\epsilon \sim \NN\big((.3,.3,.3)',.01\sigma^2_\epsilon \I\big)$, $\beta \sim \NN^+(1,.01)$, $\tau \sim \NN^+(.1,.0004)$, and $\sigma^2_\epsilon \sim IG(25,2)$.

\begin{figure}
\centering
\includegraphics[width=.9\textwidth]{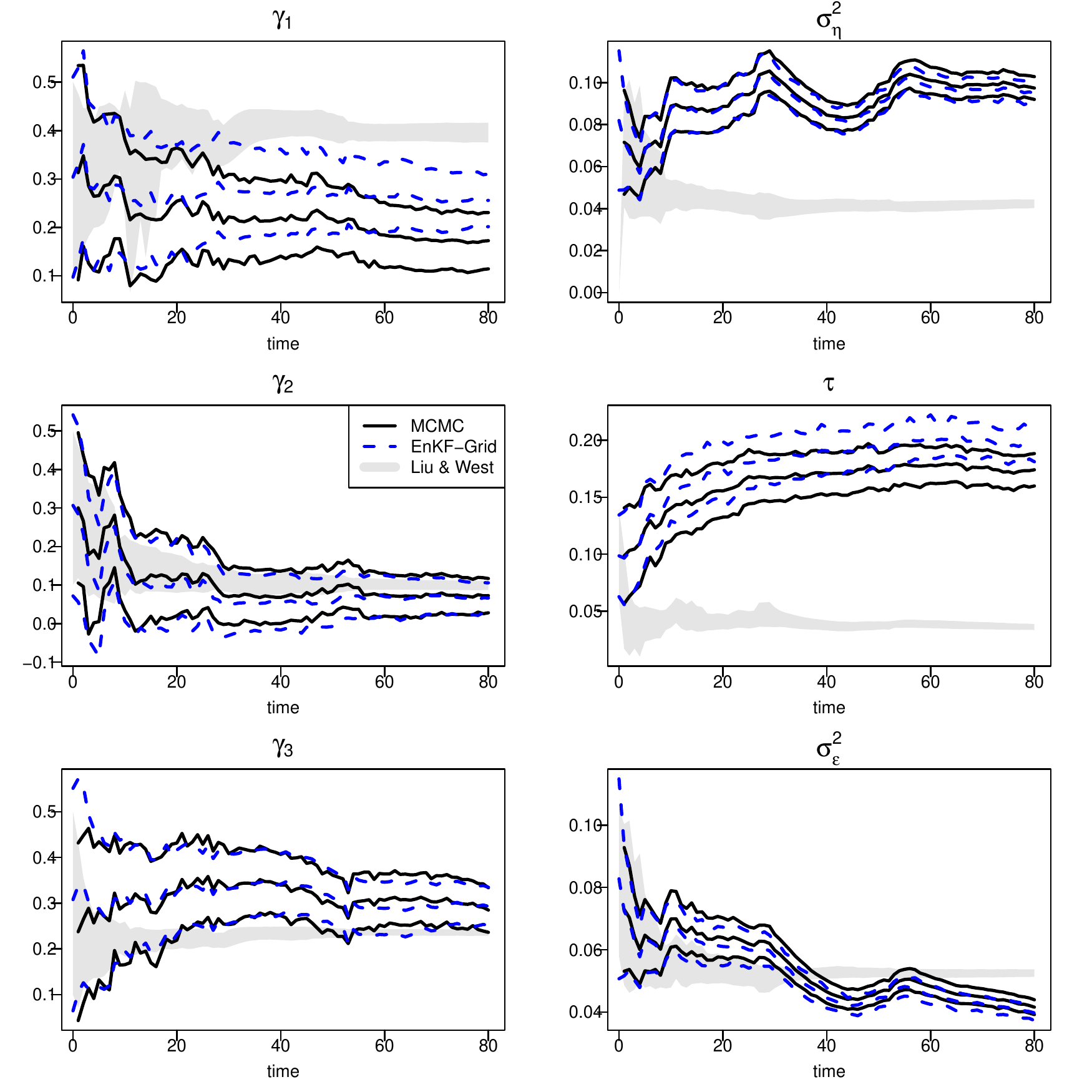}
\caption{For the cloud data in Section \ref{sec:cloud},  true posterior distributions (filtered mean and 95\% bands) of the parameters over time (in black), together with the corresponding approximations using the EnKF-Grid (blue dashed lines) and the particle filter (gray bands) of \cite{LiuWest:01}
 \label{fig:clouddata}}
\end{figure}

We applied our Algorithm \ref{alg:enkflearning} to the data using $N=100$ ensemble members and no tapering or covariance inflation. 
While the model includes six unknown parameters in total, the autoregressive parameters $\ga$ were included in the state and handled with state augmentation, and the technique of \citet{StroBeng:07} was used for inference on $\sigma^2_\epsilon$ (see Section \ref{sec:merge}). This leaves us with the parameter vector $\bth = (\tau, \beta)'$ to which we apply our method.

In Figure \ref{fig:clouddata}, the results are compared to the true posterior distribution as obtained by an MCMC procedure, and to \cite{LiuWest:01}'s particle filter with state augmentation using $N=10,000$ particles and a tuning parameter $\delta=.98$.   As we can see, the EnKF-Grid posterior does well and is close to the true posterior. The results for EnKF-Normal (not shown) are similar. Despite the very large ensemble size, the APF does not perform well, in that the means are way off, and the posterior uncertainty appears to be strongly underestimated.

\subsection{ The Lorenz 96 Model \label{sec:lorenz}}

We now consider the 40-variable system of \citet{Lore:96}, commonly referred to as the Lorenz-96 model, which mimics advection at equally-spaced locations along a latitude circle.  The
differential equations defining the time evolution of the system are given by
\begin{equation*}
\dot{x}_{t,k} = (x_{t,k+1}-x_{t,k-2})x_{t,k-1}-x_{t,k} + F,
\end{equation*}
for $k=1,\ldots,n=40$, with periodic boundary conditions.  We note that the system equations contain quadratic non-linearities which define a non-linear transition function $\mathcal{M}(\cdot)$, and also that $\Q=\zero$ ({\em cf.} equation (\ref{eq:evo})).  Here, we set the forcing parameter $F$ equal to $8$, and the time step $\delta$ equal to 0.25, resulting in a forward map with significant nonlinearities yielding distinctly non-Gaussian forecast distributions \citep[see Fig.~2 in][]{BengSnydNych:03}.  A numerical solver is used to propagate the system over time.

We simulate the true value $\x_0^*$ of the initial state from a long run of the Lorenz-96 model. At each time $\delta t, ~ t=1,2,\ldots,250$ we take $m=n$ noisy observations according to \eqref{eq:obs} with $\H = \I$. We assume spatially correlated observation errors, with $\R\equiv\R(\bth)$ defined by the Mat\'ern covariance model
$$
K(d;\theta) = \frac{\sigma^2}{2^{\nu-1}\Gamma(\nu)}
\left(\frac{d}{\lambda}\right)^\nu \mathcal{K}_\nu
\left(\frac{d}{\lambda}\right),
$$ 
where $\sigma^2$ is the sill parameter, $\lambda$ is the spatial range
parameter, $\nu$ is the smoothness parameter, and the distance $d$ between $x_{t,i}$ and $x_{t,j}$ is defined as $\min\{ |i-j|, 40-|i-j| \}$.  Data are simulated
using the parameter values $(\sigma^2, \lambda, \nu) = (1, 1, .5)$. We take the initial state distribution to be $\x_0 \sim \NN(\x_0^*,0.25\I)$, and assume the following independent prior distributions for the parameters: 
$\sigma^2 \sim \IG(5,5)$, $\lambda \sim \NN^+(1,.64)$, and $\nu \sim \NN^+(.25,.25)$.

\begin{figure}
\centering
\includegraphics[width=.95\textwidth]{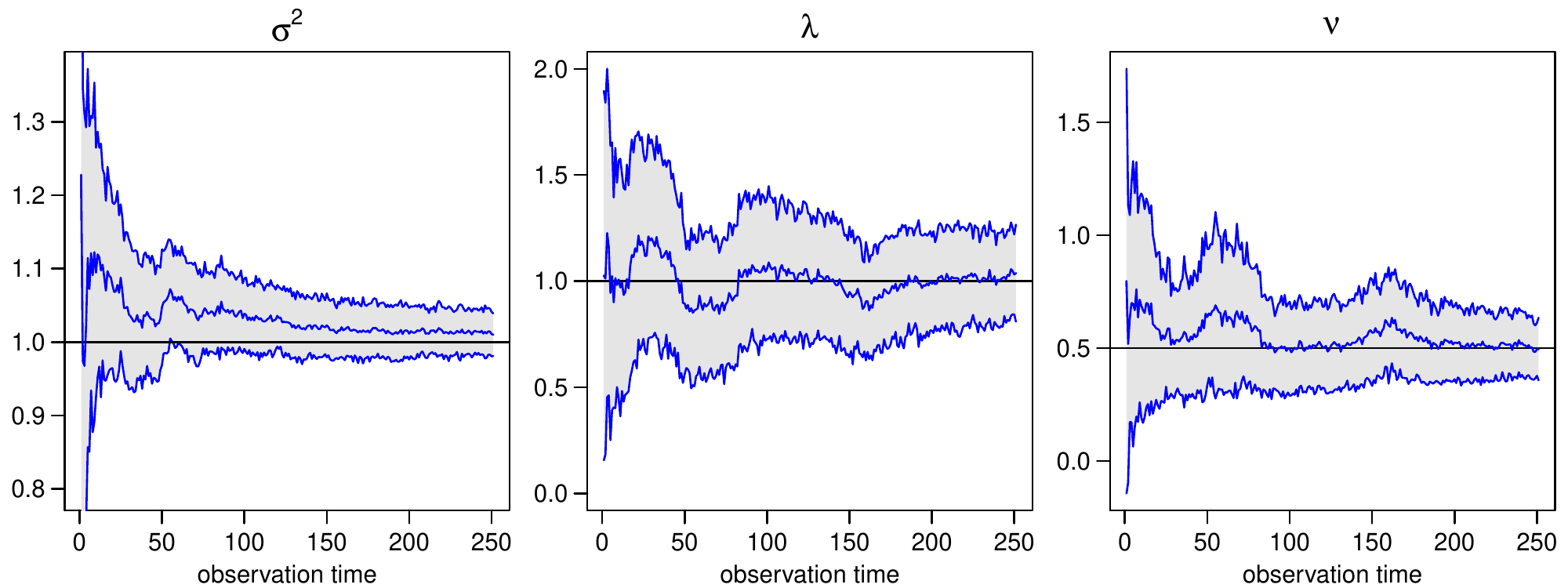}
\caption{For data simulated from the Lorenz-96 model (see Section \ref{sec:lorenz}), the marginal posterior distributions (filtered mean and 95\% bands) of the parameters over time \label{fig:lorenz}}
\end{figure}

\begin{figure}
\centering
\includegraphics[width=.8\textwidth]{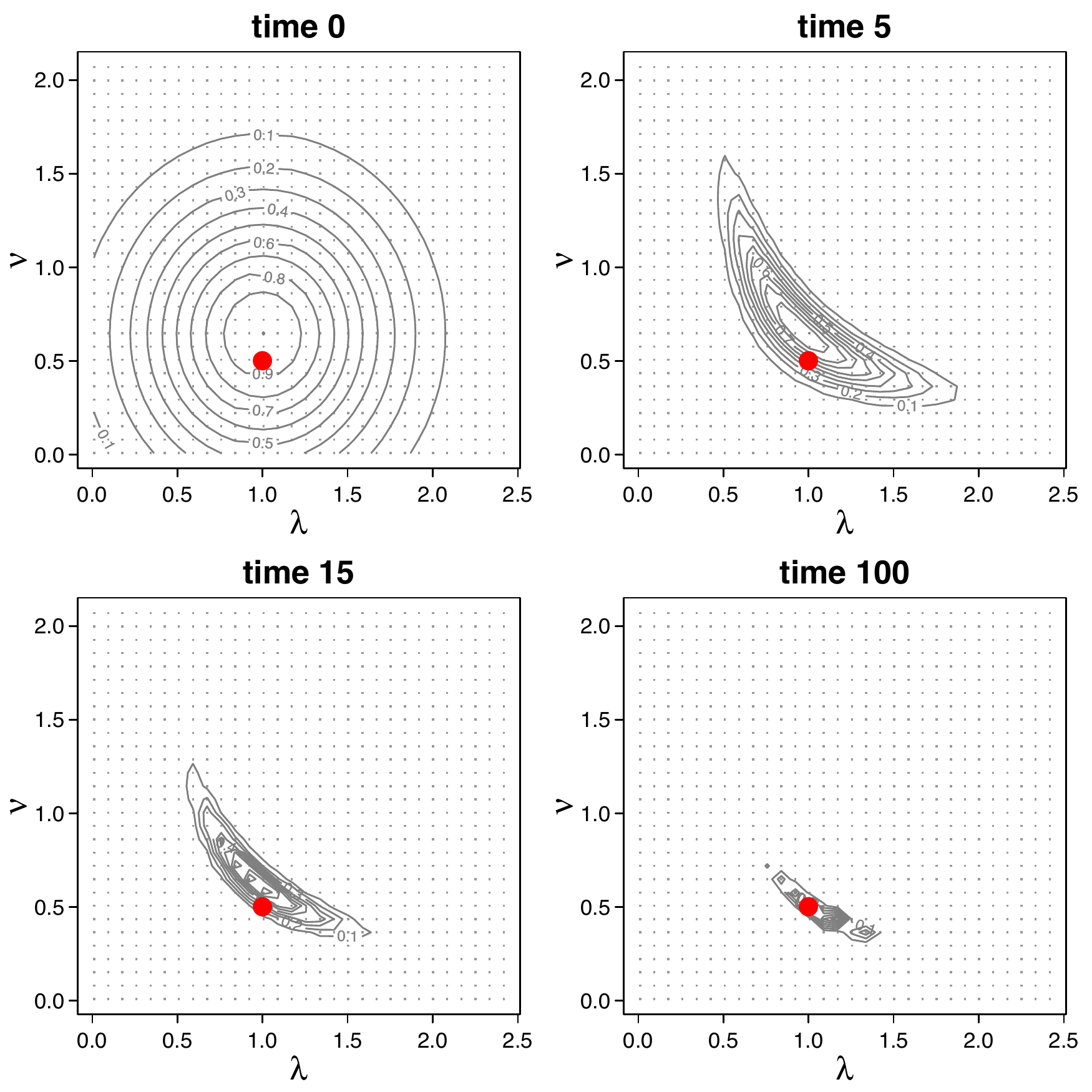}
\caption{For data simulated from the Lorenz-96 model (see Section \ref{sec:lorenz}), contours of the joint posterior distribution of the parameters $\lambda$ and $\nu$ at several time points. The contour values are normalized to yield a maximum posterior density of 1 at each time point. The true value of $(\lambda,\nu)=(1,.5)$ is indicated by a black dot. \label{fig:lorenzbanana}}
\end{figure}

Using the method of \citet{StroBeng:07} to handle $\sigma^2$, we applied our EnKF-Grid algorithm for inference on $\lambda$ and $\nu$ to the resulting simulated data with 20 grid points per parameter, $N=100$ ensemble members, and a \citet{GaspCohn:99} taper with range 12. The marginal posterior distributions of the three parameters over time are shown in Figure \ref{fig:lorenz}. As we can see, the posterior distributions again seem to be converging to the true values. The EnKF-Grid also produces estimates of the joint posterior distribution of the parameters. Figure \ref{fig:lorenzbanana} shows the strong posterior dependence between $\lambda$ and $\nu$ at several time points.

\section{Conclusions \label{sec:conclusions}}

We have presented new algorithms for sequential state and parameter estimation that combine information about the parameters from data at different time points in a consistent probabilistic framework. The algorithms obtain the marginal posterior distribution of the parameters at each time point using a grid or normal approximation, while the distribution of the states given the parameters is obtained by the EnKF. The methods can also be combined with existing approaches for parameter estimation in the EnKF, such as state augmentation. We have shown in several numerical examples that the  posterior distribution of the parameters as approximated by our methods is close to the true posterior, converges to the true parameter value, and strongly outperforms popular existing approaches.

While the current software implementation of our approach is not suitable for applications with truly high-dimensional states, we expect our methods to work in high dimensions as well, as long as the embedded EnKF is well suited and well tuned to the application if the parameters are known.

A separate question is how our method will scale to high-dimensional parameter vectors (i.e., a large number of unknown parameters). The computational cost of the EnKF-Grid approach is exponential in the number of unknown parameters, and this approach is hence most suitable when the number of parameters (minus the parameters that can be handled by state augmentation and other methods) is in the single digits. The computational cost of the EnKF-Normal approach is cubic in the number of parameters and should thus scale to moderately high parameter dimensions, although the cost of the optimization procedure to find the posterior mode at each time point might become prohibitive.

Note that our methods ``break'' the dependence between the parameter approximations at successive time points and are, in their present form, unable to approximate the joint dependence structure in the posterior distribution of the parameters at different time points. This could potentially be remedied using a shift-based update to the parameters. In the case of the normal approximation, this shift would be similar to the one used for the state update in the EnKF, while in the grid-based approximation a shift based on a piecewise linear approximation \citep[cf.][]{Ande:10} to the parameter density might be possible.


\small
\appendix
\section*{Acknowledgments} 

Katzfuss' research was partially supported by U.S.\ National Science Foundation (NSF) Grant DMS-1521676. Wikle acknowledges the support of NSF and the U.S.\ Census Bureau under NSF grant SES-1132031, funded through the NSF-Census Research Network (NCRN) program.

\bibliographystyle{apalike}
\bibliography{bibjon}

\end{document}